\documentclass[11pt]{article}
\usepackage{amsmath,amssymb}
\usepackage[dvips]{epsfig}
\usepackage[dvips]{color}
\usepackage{pb-diagram,lamsarrow,pb-lams}
\DeclareFontFamily{OT1}{rsfs}{}
\DeclareFontShape{OT1}{rsfs}{m}{n}{ <-7> rsfs5 <7-10> rsfs7 <10-> rsfs10}{}
\DeclareMathAlphabet\mathcurl{OT1}{rsfs}{m}{n}
\def\1#1{{\bf #1}}
\def\2#1{{\cal #1}}\def\9#1{{\sl #1}}\def\4#1{{\tt #1}}\def\5#1{{\sf #1}}
\def\6#1{{\mathfrak #1}}\def\7#1{{\mathbb #1}}\def\8#1{{\rm #1}}
\def\9#1{{\mathcurl #1}}\def\0#1{{\boldsymbol #1}}

\def\3{{\ss}}

\def\skb{\vskip 0.5cm}

\def\beq{\begin{eqnarray}}
\def\eeq{\end{eqnarray}}
\def\vs{\vspace{0.2cm} \\}

\def\Poin{{\8P^\uparrow_+}}

\newtheorem{The}{Theorem}[section]
\newtheorem{Def}[The]{Definiton}
\newtheorem{Lem}[The]{Lemma}
\newtheorem{Pro}[The]{Proposition}
\newtheorem{Cor}[The]{Corollary}
\def\bdef{\begin{Def}\1: \em}
\def\eef{\end{Def}}

\def\blem{\begin{Lem}\1: }
\def\elem{\end{Lem}}
\def\bthe{\begin{The}\1: }
\def\ethe{\end{The}}
\def\bpro{\begin{Pro}\1: }
\def\epro{\end{Pro}}
\def\bcor{\begin{Cor}\1: }
\def\ecor{\end{Cor}}

\def\al{\alpha}
\def\be{\beta}
\def\Gam{\Gamma}

\def\te{\theta}

\def\sgm{\sigma}
\def\om{\omega}\def\Om{\Omega}

\def\bpr{\paragraph*{\it Proof.}}

\def\epr{$\square$\skb}

\def\<{\langle}
\def\>{\rangle}
\def\Ad{{\rm Ad}}

\def\bdes{\begin{enumerate}}
\def\edes{\end{enumerate}}

\def\bmat{\left( \begin{array}{ccc} }
\def\emat{\end{array} \right)}
\def\bcase{\left\{ \begin{array}{ccc} }
\def\ecase{\end{array} \right\}}

\def\beqa{\begin{eqnarray*}}
\def\eeqa{\end{eqnarray*}}
\def\bdia{\begin{diagram}}
\def\edia{\end{diagram}}

\title{\bf Application of Tomita-Takesaki theory
in algebraic euclidean field theories}
\author{{\it Dirk Schlingemann} \\
The Erwin Schr\"odinger International Institute \\ 
for Mathematical Physics (ESI)\\
Vienna}
\begin{document}
\maketitle
\abstract{
The construction of the known interacting  quantum field theory models
is mostly based on euclidean techniques.
The expectation values of interesting 
quantities are usually given in terms of 
euclidean correlation functions from which one 
should be able to extract information about the
behavior of the correlation functions of the Minkowskian counterpart.

We think that the C*-algebraic approach to euclidean 
field theory gives an appropriate setup in order to study structural aspects
model independently. 
A previous paper deals
with a construction scheme which relates to each 
euclidean field theory a Poincar\'e covariant quantum field theory 
model in the sense of R. Haag and D. Kastler. 

Within the framework of R. Haag and D. Kastler,
the physical concept of PCT symmetry and spin and statistics 
is related to the Tomita-Takesaki theory of von Neumann algebras
and this important aspects has been studied by several authors. 

We express the PCT symmetry 
in terms of euclidean reflexions and we explicitly identify the 
corresponding modular operator and the modular conjugation 
of the related Tomita-Takesaki theory.  
Locality, wedge duality, and a geometric action of the modular group 
of the von Neumann algebra of observables, localized within a 
wedge region in Minkowski space, are direct consequences.}

\newpage
\section{Introduction}
\label{sc1}
Concerning the know examples for non-free quantum field theory models,
the construction of them, by means of euclidean techniques, is the 
most successful method which is known.
Not surprisingly, most of the interesting 
quantities of such a model are explicitly given only in terms of 
euclidean correlation functions. 
It is therefore natural to ask the following question:

Given a euclidean field theory from which a quantum field theory 
can be constructed. Which properties of the quantum field theory 
can directly read off from the euclidean data?

This motivates the development of 
tools which analyze structural aspects of 
euclidean field theories in a systematic manner and  
we think that the C*-algebraic approach of euclidean 
field theory gives an appropriate setup in order to follow this 
program. Analogously to the famous Osterwalder-Schrader 
Theorem \cite{OstSchra1}, it can be shown \cite{Schl97} that to each 
euclidean field theory, formulated in the C*-algebraic 
framework, a quantum field theory 
model in the sense of R. Haag and D. Kastler \cite{H,HK}
can be associated. We give a brief description of the corresponding 
construction scheme later.

Based on this work, a tool for investigating the high energy 
behavior of a quantum field theory model, by only looking
at its euclidean counterpart, has already been discussed
in \cite{Schl99a}. Compared to the 
scaling limit analysis of D. Buchholz and R. Verch
\cite{Bu97,BuVer97,Bu96a,Bu96b,BuVer95} one 
finds the expected results, namely that the 
high energy behavior of the euclidean model reflects 
the high energy behavior of the corresponding quantum field theory.

In addition to that, the C*-algebraic point of view provides new strategies 
for constructing euclidean field theory models, as 
it is laid out in \cite{Schl98,Schl99c}.

Within the framework of algebraic quantum field theory,
the physical concept of PCT symmetry and spin and statistics 
is deeply linked to the Tomita-Takesaki theory of von Neumann algebras
and has been studied in several papers, see for example  
\cite{BiWi75,Bor95a,Bor95b,GuLo95,Kuck95,Yng96,Sum96,Bor98,BorYng99}.
Moreover, Tomita-Takesaki theory might also be important 
for constructive purposes as it is, for example,  
proposed in \cite{SchrWies98}.

From the point of view of eulidean field theory, it would be 
desirable to express the modular data of a wedge 
algebra of the quantum field theoretical counterpart 
directly in terms of euclidean correlation functions. 
This might be of importance, since
according to the discussion in \cite{SchrWies98}, 
there is hope, that an analysis of the modular data related to the 
euclidean field theory 
models, which are constructed in \cite{Schl98,Schl99c} by an 
abstract procedure, can be used as a tool in order to 
decide whether a theory describes interaction.

\paragraph{\em Organization of the paper.}
In the second part of the present section, we 
make some preliminary remarks on the algebraic approach to
euclidean field theory in order to introduce notations and conventions 
which are used.
Starting from a given euclidean field, we present in Section \ref{sc2} 
the main results which state in particular that 
the square-root of the modular operator of a
wedge algebra of the quantum field theory, constructed from 
the euclidean data, can explicitly be identified with a 
particular euclidean rotation with rotation angle $s=\pi$. 
Furthermore, the corresponding modular conjugation is 
a PCT symmetry which can be expressed in terms of euclidean reflexions.
We close the paper by Section
\ref{seccr}, mentioning some work in progress.
We feel obliged to postpone most of the technical details and the proofs 
to the appendix in order to keep the paper more readable.

\paragraph{\em Preliminary remarks on the algebraic approach to
euclidean field theory.}
The starting point within the C*-algebraic approach to 
euclidean field theory is a so called
{\em euclidean net of C*-algebras} $(\6B,\be)$. Such a net is an
inclusion preserving prescription 
$\9U\mapsto\6B(\9U)\subset \6B$, 
which assigns to each bounded convex region $\9U$ in $\7R^d$
a C*-algebra $\6B(\9U)$. This assignment has to fulfill 
several assumptions, according to physical principles.

Two operators commute if they are localized in 
disjoint regions, more specifically,
if the intersection $\9U_1\cap\9U_2=\emptyset$ is empty, then
the commutator $[b_1,b_2]=0$ vanishes 
for all operators $b_1\in\6B(\9U_1)$ and $B_2\in\6B(\9U_2)$.
The net $\9U\mapsto\6B(\9U)$ is euclidean covariant, i.e. 
there exists a group homomorphism $\be$ form the euclidean group into the 
automorphism group of $\6B$
such that for a euclidean transformation $h\in\8E(d)$ 
the algebra of a local region $\9U$ is 
mapped, via $\be_h$, onto the algebra of the 
transformed region $h\9U$: $\be_h\6B(\9U)=\6B(h\9U)$. 

We have to mention that, 
although the analogy between 
the C*-algebraic approach to euclidean field theory
and the Haag-Kastler framework for quantum field theory is obious,  
the euclidean C*-algebras 
have to be interpreted in a different manner.
Within the  Haag-Kastler program, 
the dynamics of a given quantum field theory is usually 
contained in the relative inclusion of local 
algebras, whereas a euclidean net of C*-algebras only 
describes kinematical aspects.
The dynamics of euclidean field theory model is encoded 
in the choice of a particular (euclidean invariant and reflexion 
positive regular) state on the corresponding euclidean C*-algebra: 

\paragraph{\it Euclidean invariance:}
A state $\eta$ on $\6B$ is called {\em euclidean invariant}
if for each $h\in \8E(d)$ the identity 
$\eta\circ\be_h=\eta$ holds true.

\paragraph{\it Reflexion positivity:}
A state $\eta$ on $\6B$ is called {\em reflexion positive} if  
exists a euclidean direction $e\in S^{d-1}$   
such that the sesquilinear form
\beqa
\6B(\8H_e)\otimes\6B(\8H_e)\ni b_0\otimes b_1\mapsto 
\<\eta,j_e(b_0)b_1\>
\eeqa
is positive semi definite. Here, 
$\8H_e$ is the half space $\7R_+e+e^\perp$ with respect to a given 
euclidean direction $e\in S^{d-1}$, $j_e$ is 
the anti-linear involution which is given by
$j_e(b)=\be_{\te_e}(b^*)$ where $\te_e$ is the 
euclidean reflexion $\te_ex=x-2\<e,x\>e$.

\paragraph{\it Regularity:}
A state $\eta$ on $\6B$ is called {\em regular} if for each 
$b_0,b_1,b_2\in\6B$ the map
\beqa
h&\mapsto&<\eta,b_0\be_h(b_1)b_2>
\eeqa
is continuous. 
\skb

A triple $(\6B,\be,\eta)$ consisting of a euclidean net of 
C*-algebras $(\6B,\be)$ and a euclidean invariant reflexion positive
regular state $\eta$ is called a {\em euclidean field}.

\paragraph{\em From euclidean field theory to quantum field theory.}
We briefly describe here, as it has carried out in \cite{Schl97},
how to construct from 
a given euclidean field a quantum field theory. 

\paragraph{\it Step 1:}
According to the reflexion positivity, for e given direction
$e\in S^{d-1}$ there exists a Hilbert space $\2H$ and a
linear map
\beqa
\Psi:\6B(\8H_e)\mapsto \2H 
\eeqa
which is uniquely determined by 
\beqa
\<\Psi[b_0],\Psi[b_1]\>&=&<\eta,j_e(b_0)b_1>
\eeqa
and we have a distinguished vector $\Om:=\Psi[\11]$, the vacuum vector.
Following the analysis presented in \cite{FrohOstSeil}, 
a unitary strongly continuous representation of the Poincar\' e group 
$U$ on $\2H$ can be constructed (see also \cite{Seil82,Schl97}). 
The vacuum vector $\Om$ is invariant under the action of $U$ and, 
in addition to that, the spectrum of the the generator of the 
translations $x\mapsto U(x)$ is contained in the 
closed forward light cone $\bar V_+=\{x\in\7R^{1,d-1}|x^2\geq 0 ; x^0\geq 0\}$.
A more detailed description of the construction of $U$ 
is given in Appendix \ref{app1}.
 
\paragraph{\it Step 2:}
For a subset $\9V$ of the hyperplane $e^\perp$, perpendicular to $e$, 
we introduce the algebra $\6B(\9V)$ of {\em time zero opeartors},
localized in $\9V$, by the intersection
\beqa
\6B(\9V)&:=&\bigcap_{s\in\7R_+} \6B([0,s)e\times\9V) \ \ .
\eeqa
The algebar $\6B(e^\perp)$ is then the C*-algebra which is generated 
by all local time-zero algebras $\6B(\9V)$ with $\9V\subset e^\perp$.

There exists a *-representation $\pi$ of the time-zero algebra 
$\6B(e^\perp)$ on $\2H$,
which is uniquely determined by the relation
\beqa
\pi(b)\Psi[b_1]&=&\Psi[bb_1]
\eeqa
for each $b_1\in\6B(\8H_e)$. 

The algebra $\pi(\6B(e^\perp))$  
can be regarded as the {\em Cauchy data} of the 
quantum field theory model by identifying 
the hyperplane $e^\perp$ with the spacelike 
hyperplane $x^0=0$ in Minkowski space $\7R^{1,d-1}$.

For a double cone $\9O$ (a causally complete bounded set
in Minkowski space), we define $\6A(\9O)$ to be the 
von Neumann algbera on $\2H$ which is generated by 
all operators 
\beq\label{opm1}
\Pi[g,b] &:=& U(g)\pi(b)U(g)^*
\eeq
with $b\in\6B(\9V)$, such that  
$g\9V\subset\9O$ for the  
Poincar\'e transformation $g$.
The prescription 
\beqa
\9O\mapsto\6A(\9O)
\eeqa 
is an isotonous net of C*-algebras. We define a group 
homomorphism $\al$ from the Poincar\'e group 
to the automorphism group of $\6A$ by 
\beqa
\al_g:=\Ad(U(g))
\eeqa
which is, by construction, covariant, i.e.
$\al_g\6A(\9O)=\6A(g\9O)$ for each double cone $\9O$ and for each
Poincar\'e transformation $g$.

The state $\om$ on $\6A$, which is given by 
\beqa
<\om,a>&=&\<\Om,a\Om\>
\eeqa
has the following properties:

\paragraph{\it Poincar\'e invariance:}
For each Poincar\'e transformation $g$ the identity
$\om\circ\al_g=\om$ holds true which is 
a consequence of the invariance of $\Om$ under 
$U(g)$. 

\paragraph{\it Positivity of the energy:}
For each $a_1,a_2\in\6A$ and for each 
test function $f\in S(\7R^{1,d-1})$ on Minkowski space
whose Fourier transform $\tilde f$ has support in the 
complement of the closed forward light cone $\bar V_+$,
the identity 
\beqa
\int \8dx \ f(x) \ <\om,a_1\al_x a_2> &=& 0
\eeqa
is valid. This is nothing else but 
expressing the fact that the generators of the 
translations $x\mapsto U(x)$ have joint spectrum in the closed 
forward light cone.

\paragraph{\it Locality:}
If $\9O_1$ and $\9O_2$ are spacelike separated regions, then
the commutator $[a_1,a_2]=0$ vanishes 
for $a_1\in\6A(\9O_1)$ and $a_2\in\6A(\9O_2)$.

\paragraph{\em Remark.}
Whereas Poincar\'e invariance as well as 
the positivity of the energy follow directly 
from the construction of the representation $U$,
the fact that locality is fulfilled is not directly 
visible. The proof, carried out in \cite{Schl97}, is quite lengthy.
We shall see later, that the use of Tomita-Takesaki theory for 
wedge algebras, leads to a more elegant proof of locality.

\section{Modular data for wedge algebras as geometric operations in 
euclidean space}
\label{sc2}
The present section is concerned with the discussion of 
the Tomita-Takesaki theory of wedge algebras 
of the net $\9O\mapsto\6A(\9O)$ of von Neumann algebras emerging from
a given euclidean field theory model. 
We present here the main results of the paper, which relate 
the modular operator and the modular conjugation of a 
particular wedge algebra to geometric operations in 
euclidean space. The technical details and the proofs 
are postponed to the appendix.

\subsection{KMS states associated with wedge algebras}
For any wedge region $\9W$ in Minkowski space 
the wedge algebra with respect to $\9W$ is the von Neumann 
algebra 
\beqa
\6A(\9W)&:=&
\bigvee_{\9O\subset\9W}\6A(\9O) 
\eeqa
generated by all local von Neumann 
algebras $\6A(\9O)$ with $\9O\subset \9W$. 

Particular wedge algebras are related to euclidean directions 
$e_0\in S^{d-1}$ which are perpendicular $e_0\perp e$ 
to $e$. An application of
the boosts in $(e,e_0)$ direction to the half hyperplane 
$\8P_{(e,e_0)}:=\8H_e\cap\8H_{-e}\cap\8H_{e_0}$
yields the wedge region 
\beqa
\9W_{(e,e_0)}&:=&\bigcup_{t\in\7R}\exp(tB_{(e,e_0)})\8P_{(e,e_0)} \ \ .
\eeqa
Obviously, the spacelike complement $\9W_{(e,e_0)}'$ of 
the wedge $\9W_{(e,e_0)}$ is just 
the wedge $\9W_{(e,-e_0)}$ with respect to the reflected direction $-e_0$.
Writing $\al_{(e,e_0)}:\7R\to\8{Aut}(\6A)$ 
for the one-parameter automorphism group of boosts
\beqa
\al_{(e,e_0,t)}&:=&\8{Ad}\biggl[U_{(e,\eta)}( \exp(tB_{(e,e_0)}) )\biggr]
\eeqa
we obtain a W*-dynamical system 
$(\6A(\9W_{(e,e_0)}),\al_{(e,e_0)})$ together with a
$\al_{(e,e_0)}$-invariant state 
\beqa
\om_{(e,e_0)}&:=& \om|_{\6A(\9W_{(e,e_0)})} \ \ ,
\eeqa
the restriction of the vacuum state $\om$ to the 
corresponding wedge algebra. 

\bthe\label{the0}
For each direction $e_0\perp e$, the 
state $\om_{(e,e_0)}$ is a KMS state at inverse 
temperature $\be=2\pi$ with respect to the W*-dynamical system 
$$(\6A(\9W_{(e,e_0)}),\al_{(e,e_0)}) \ \ . $$
\ethe

The proof of Theorem \ref{the0} can be obtained in complete analogy 
to the analysis of \cite{KlLan81} and we give a 
version of it within the Appendix \ref{app2}. 

\subsection{The PCT symmetry and complex Lorentz boosts as
geometric operations in euclidean space}
\paragraph{\em The $e_0$-PCT operator.}
For a euclidean direction $e_0\in S^{d-1}$ which is perpendicular to $e$,  
the euclidean reflexion $$\te_{e_0}:x\mapsto x-2\<e_0,x\>e$$ 
is contained in the stabilizer group of $e$ and hence it 
gives rise to an anti-unitary operator
$\9J_{(e,e_0)}$, called the  $e_0$-PCT operator. 
It is defined according to the prescription  
\beqa
\9J_{(e,e_0)}\Psi[b]
&:=&
\Psi[j_{e_0}(b)] 
\eeqa 
and it has the geometric property (see Appendix \ref{app3} for the proof):

\bpro\label{pro1}
For each $e_0\in S^{d-1}\cap\8P_e$ the identity 
\beqa
\9J_{(e,e_0)}\6A(\9W_{(e,e_0)})\9J_{(e,e_0)}
&=&\6A(\9W_{(e,-e_0)}) 
\eeqa
is valid.
\epro
  
\paragraph{\em Complex Lorentz boosts as euclidean rotations.}
For a direction $e_0\in S^{d-1}$, $e\perp e_0$, 
The one-parameter group $t\mapsto U( \exp(tB_{(e,e_0)})$
can be extended analytically to complex parameters
on a appropriate dense subspace of $\2H$. 
This is based on  remarkable facts which have 
been established by J. Fr\"ohlich 
\cite{Froh80}, on one hand, and by A. Klein and L. J. Landau \cite{KlLan81a},
on the other hand.

The generator $L_{(e,e_0)}\in\6so(d)$ 
of the euclidean rotations within the $e-e_0$ plane 
yields a one-parameter group of automorphisms on $\6A$  
\beqa
\be_{(e,e_0,s)}&:=&\be_{\exp(sL_{(e,e_0)})}  \ \ .
\eeqa
For each $s\in (-\pi/2,\pi/2)$ an operator 
$V_{(e,e_0)}(s)$ is uniquely determined by 
\beqa
V_{(e,e_0)}(s)\Psi[b]
&=&\Psi[\be_{(e_1,e,s)}b] 
\eeqa
for each $b\in\6B(\8H_e)$ with $\be_{(e,e_0,s)}b\in\6B(\8H_e)$.
On an appropriate dense subspace $\2D\subset\2H$,
the operator $V_{(e,e_0)}(s)$ is related to 
the one-parameter group $t\mapsto U( \exp(tB_{(e,e_0)}))$ by
\beqa
V_{(e,e_0)}(s)&=&U( \  \exp(\8isB_{(e,e_0)} ) \ ) \  \ .
\eeqa

\subsection{The modular operator and the modular conjugation
for wedge algebras.}
From the fact that $\om_{(e,e_0)}$ is a KMS state for the 
W*-dynamical system 
$(\6A(\9W_{(e,e_0)}),\al_{(e,e_0)})$, one concludes 
(e.g. \cite{Araki76})
that the vector $\Om$ is separating for 
the algebra $\6A(\9W_{(e,e_0)})$.

Therefore there exists a modular operator $\Delta_{(e,e_0)}$
and a modular conjugation $J_{(e,e_0)}$ with respect to the 
pair $(\6A(\9W_{(e,e_0)}),\Om)$. 
The subsequent theorem, which is proven in the Appendix \ref{app4},
states that the modular data can be expressed in terms of 
geometric actions in {\em euclidean} space:

\bthe\label{the1}
For the modular operator $\Delta_{(e,e_0)}$ and the modular 
conjugation $J_{(e,e_0)}$ the identities 
\beqa
\Delta_{(e,e_0)}^{1/2}&=&V_{(e,e_0)}(\pi)
\vs\vs
J_{(e,e_0)}&=&\9J_{(e,e_0)}
\eeqa
hold true for each $e_0\in S^{d-1}\cap \8P_e$.
\ethe

\paragraph{\em Locality and wedge duality.} 
The statement of Theorem \ref{the1} has some direct implications.
As already mentioned, a proof for locality of the net 
$\9O\mapsto\6A(\9O)$ has already been established in 
\cite{Schl97}. The proof of Theorem \ref{the1},
which we postpone to the appendix, do not make use of this fact
and by means of Theorem \ref{the1} we derive 
an independent proof for locality which is, compared to  \cite{Schl97}, 
much more elegant and straight forward.

\bcor
The net $\9O\mapsto \6A(\9O)$ fulfills wedge duality,
i.e. for each wedge region $\9W$ the identity
\beqa
\6A(\9W)'=\6A(\9W')
\eeqa
is valid. In particular, locality holds true.
\ecor
\bpr
As a consequence of Theorem \ref{the1} we get 
\beqa
\6A(\9W_{(e,e_0)})'&=&\6A(\9W_{(e,-e_0)}) 
\eeqa
and the Poincar\'e covariance 
of the net $\9O\mapsto\6A(\9O)$ implies wedge duality
for each wedge $\9W$. Therefore, the net fulfills locality since 
for two spacelike separated double cones $\9O_1\subset \9O_2'$
there is a wedge $\9W$ with 
$\9O_1\in \9W$ and $\9O_2\in \9W'$.  Now wedge duality implies 
for operators $a_1\in\6A(\9O_1)$ and 
$a_2\in\6A(\9O_2)$ that $[a_1,a_2]=0$.
\epr

\section{Concluding remarks}
\label{seccr}
We have shown, that the modular operator and the modular conjugation 
of a particular wedge algebra $\6A(\9W)$, associated to 
a given euclidean field $(\6B,\be,\eta)$, 
do not only have the meaning as geometric action on 
Minkowski space in terms of Lorentz boosts and reflexions,
they also can be identified with geometric operations 
in euclidean space, namely particular euclidean rotations and 
euclidean reflexions. 
This fact can be used to conclude wedge duality 
for the net $\9O\mapsto\6A(\9O)$. 
As a consequence we get, compared to the analysis 
carried out in \cite{Schl97},  an improved method in order 
to prove locality. 
 
Keeping in mind that the minkowskian analogue of the euclidean 
$d$-sphere $S^d\subset\7R^{d+1}$ is the de Sitter space, 
it should be possible, by exploring the analytic structure of de Sitter space, 
to construct from a given euclidean field theory 
$(\6B,\be,\eta)$ on the sphere $S^d$ 
a quantum field theory $(\6A,\al,\om)$
in de Sitter space (A forthcoming preprint is in preparation). 
For an example, we refer the reader to the 
work of R. Figari, R. H\"oegh-Krohn, and C. R. Nappi
\cite{FigHoehNap75}.
According to Theorem \ref{the0},
we conjecture that the reconstructed state $\om$
fulfills the so called {\em geodesic KMS condition}, 
in the sense of H. J. Borchers 
and D. Buchholz \cite{BorBu98}, 
i.e. for any geodesic observer the state $\om$ looks like 
an equilibrium state.
In order to prove locality for the constructed quantum field 
theory in de Sitter space, the method of \cite{Schl97},
can not directly be applied since here the euclidean translations 
are used. Establishing the analogous results of Theorem \ref{the1}
for the theory in de Sitter space, locality would also follow here. 

\subsubsection*{{\it Acknowledgment:}}
I am grateful to Prof. Jakob Yngvason for 
supporting this investigation many ideas.
This investigation is financially supported by the 
{\em Jubil\"aumsfonds der Oesterreichischen Nationalbank} 
which is also gratefully acknowledged.
Finally I would like to thank the 
Erwin Schr\"odinger International Institute for Mathematical Physics, 
Vienna (ESI) for its hospitality.

\newpage
\begin{appendix}
\section{Construction of a representation of the Poincar\'e group}
\label{app1}
In order to keep the paper self contained, we review here the 
construction procedure of a representation of the Poincar\'e group
from a given eucliden field $(\6B,\be,\eta)$ in more ditail. 

We first  give a list of facts which are consequences of the 
axioms for a euclidean field.

\paragraph{\em Fact 1:} 
For each $e\in S^d$ there is a strongly continuous 
one-parameter semi-group of contractions $T_e$
with positive generator $H_e\geq 0$ which is given according to 
\beqa
T_e(s)\Psi[b]&=&\Psi[\be_{se}b] 
\eeqa
for $s>0$ and for $b\in\6B(\8H_e)$.

\paragraph{\em Fact 2:}
For $e\in S^{d-1}$, $w\in (0,\pi/2)$, the conic 
region $\Gam(e,w)$ is defined 
to be the $\8O_e(d-1)$ invariant 
cone in $e$ direction with opening angle $w$.
Moreover, for a pair euclidean time 
directions $e,e_1\in S^{d-1}$, $e\perp e_1$,
the generator of the rotations within the $(e,e_1)$ plane is denoted by
$L_{(e,e_1)}\in\6so(d)$.
Hence the corresponding one-parameter group yields a one-parameter group of 
automorphisms on $\6B$  
\beqa
\be_{(e,e_1,s)}&:=&\be_{\exp(sL_{(e,e_1)})} 
\eeqa
and a local one-parameter group 
\beqa
V_{(e,e_1)}=\{V_{(e,e_1)}(s)|s\in (-\pi/2,\pi/2)\} 
\eeqa
of selfadjoint operators. For each 
$s\in (-\pi/2,\pi/2)$ the domain of $V_{(e,e_1)}(s)$ is 
\beqa
\2D(e,|s|) &:=&\Psi[\6B(\Gam(e,\pi/2-|s|))] 
\eeqa
for each $s\in (-\pi/2,\pi/2)$.
The operator $V_{(e,e_1)}(s)$ is uniquely determined by 
\beqa
V_{(e,e_1)}(s)\Psi[b]
&=&\Psi[\be_{(e_1,e,s)}b] 
\eeqa
and there exists an anti-selfadjoint operator $\1B_{(e,e_1)}$ on 
$\2H$ such that
\beqa
V_{(e,e_1)}(s)&=&\exp(\8is\1B_{(e,e_1)}) \  \ .
\eeqa
For each angle $w\in(0,\pi/2)$,
the vectors in $\2D(e,w)$ are analytic for $B_{(e_1,e)}$.
This remarkable facts have been established by J. Fr\"ohlich 
\cite{Froh80}, on one hand, and by A. Klein and L. J. Landau \cite{KlLan81a},
on the other hand.

\paragraph{\em Fact 3:}
There is a unitary strongly continuous representation $W$ 
on $\2H$ of the stabilizer subgroup $\8E_e(d-1)$ fulfilling 
\beqa
W(g)\Psi[b]&=&\Psi[\be_g b] 
\eeqa
for $g\in\8E_e(d-1)$ and for $b\in\6B(\8H_e)$.

\paragraph{\em A representation of the Poincar\'e group.}
By making use of the analysis of virtual representations 
\cite{FrohOstSeil} a strongly continuous unitary representation
\beqa
U\in\8{Hom}[\Poin,\8U(\2H)]
\eeqa
of the Poincar\'e group can be constructed. 
The paper \cite{FrohOstSeil} exploits the facts, listed above,
and proceeds in several steps: 
 
\paragraph{\em Step 1:}
According to {\it Fact 3} the stabilizer subgroup 
$\8E_e(d-1)\subset \8E(d)$ is represented by $W$ and we put 
\beqa
U(g)&:=&W(g)
\eeqa
for each $g\in\8E_e(d-1)$. 

\paragraph{\em Step 2:}
By using {\it Fact 2} for each $e_1\perp e$ 
there exists a anti-selfadjoint 
operator $B_{(e,e_1)}$ on $\2H_{(\eta,e)}$ 
such that 
\beqa
V_{(e,e_1)}(s)&=&\exp(-\8i s \1B_{(e,e_1)}) \ \ .
\eeqa
Let $B_{(e,e_1)}\in\6o(d-1,1)$ be the boost generator 
in $e_1-e$ direction we define a unitary operator 
by 
\beqa
U(\exp(tB_{(e_1,e)}))&:=&\exp( t \1B_{(e_1,e)}) \ \ .
\eeqa

\paragraph{\em Step 3:}
Finally, the translations with respect to the time-like direction 
which corresponds to the $e$-direction in euclidean space are 
represented by 
\beqa
U(te)&:=& \exp(\8i t H_e)  
\eeqa
according to {\it Fact 1}.

\section{Proof of Theorem \ref{the0}}
\label{app2}
The main steps of the proof can be performed in complete analogy to the 
the analysis of \cite{KlLan81}. 
We consider a family of operators $b_1,\cdots, b_n$ which are contained 
in the time slice algebra, where 
$b_j\in\6B(\9V_j)$ is localized 
in a convex subset $\9V_j\subset\8H_{(e,e_0)}$. This implies that 
$\be_{(e,e_0,s)}b_j\in\6B(\8H_e)$ for each $s\in(0,\pi)$ where
we have chosen the condition $\exp(2\pi L_{(e,e_0)})=1$. 
We introduce the open subset in $\7R^2$
\beqa
I(\9V_j)&:=&\{(\tau,s)\in\7R^2|\forall e_2:
\exp(sL_{(e,e_0)})\exp(\tau L_{(e,e_2)})\9V_j\subset\8{H}_e\}
\eeqa
which contains in particular the set $\{0\}\times(0,\pi)\subset I(\9V_j)$.
We put $V:=V_{(e,e_0)}$ and $\Om=\Psi[\11]$.
By introducing the operators 
\beqa
\1b_j(\tau_j)&:=&V_{(e,e_j)}(\tau_j)\pi(b_j)V_{(e,e_j)}(-\tau_j)
\vs
&\mbox{ and }&
\vs
b_j(\tau_j,s_j)&:=&\be_{(e,e_0,s_j)}\be_{(e,e_j,\tau_j)}b_j  
\eeqa
we obtain 
\beqa
&&V(s_k)\1b_k(\tau_k)\cdots V(s_1)\1b_1(\tau_1)\Om
\vs\vs
&=&
V(s_k)\1b_k(\tau_k)V(-s_k)V(s_k+s_{k-1})\1b_{k-1}(\tau_{k-1})
V(-s_k-s_{k-1})\cdots
\vs\vs
&&\cdots V(s_1)\1b_1(\tau_1)\Om
\vs\vs
&=&
V(s_k)\1b_k(\tau_k)V(-s_k)\cdots V(s_1+\cdots+s_k)\1b_1(\tau_1)
\Om
\vs\vs
&=&
\Psi[b_k(\tau_k,s_k)\cdots b_1(\tau_1,s_1+\cdots+s_k)] \ \ .
\eeqa
We compute for $s_1,\cdots,s_n\in\7R_+$,  
$(\tau_j,s_n+\cdots+s_j)\in I(\9V_j)$ for $n\leq j\leq k+1$ and
$(\tau_i,s_k+\cdots+s_i)\in I(\9V_i)$ for $k\leq i\leq 1$:
\beqa
&&\<V(s_n)\1b_n(\tau_n)\cdots V(s_{k+1})\1b_{k+1}(\tau_{k+1})\Om,
V(s_k)\1b_k(\tau_k)\cdots 
\vs\vs
&&\cdots
V(s_1)\1b_1(\tau_1)\Om\>
\vs\vs
&=&
<\eta,j_e[ b_n(\tau_n,s_n)\cdots b_{k+1}(\tau_{k+1},s_n+\cdots +s_{k+1}) ]
\vs\vs
&&\times  
b_k(\tau_k,s_k)\cdots b_{1}(\tau_1,s_k+\cdots +s_1)> 
\vs\vs
&=&
<\eta, b_{k+1}^*(-\tau_{k+1},-s_n-\cdots -s_{k+1})\cdots b_n^*(-\tau_n,-s_n)
\vs\vs
&&\times \
b_k(\tau_k,s_k)\cdots b_{1}(\tau_1,s_k+\cdots +s_1)> \ \ .
\eeqa
Furthermore, we find due to our periodicity condition
$\exp(2\pi L)=1$:
\beqa
&&
<\eta, b_{k+1}^*(-\tau_{k+1},-s_n-\cdots -s_{k+1})\cdots b_n^*(-\tau_n,-s_n)
\vs\vs
&&\times \ 
b_k(\tau_k,s_k)\cdots b_{1}(\tau_1,s_k+\cdots +s_1)> 
\vs\vs
&=&
<\eta, b_{k+1}^*(-\tau_{k+1},-\pi-s_n-\cdots -s_{k+1})\cdots 
b_n^*(-\tau_n,-\pi-s_n)
\vs\vs
&&\times \ \
b_k(\tau_k,-\pi+s_k)\cdots b_{1}(\tau_1,-\pi+s_k+\cdots +s_1)> 
\vs\vs
&=&
<\eta, b_k(\tau_k,-\pi+s_k)\cdots b_{1}(\tau_1,-\pi+s_k+\cdots +s_1)
\vs\vs
&&\times \ \
b_{k+1}^*(-\tau_{k+1},-(-\pi+s_n+\cdots +s_{k+1}))\cdots 
b_n^*(-\tau_n,-(-\pi+s_n))>  \ \ .
\eeqa
In the last step we have used the 
locality of the euclidean  net $(\6B,\be)$, i.e.  
operators which are localized in disjoint regions commute.
According to the definition of $I(\9V_j)$ we have for  
$(\tau,s)\in I(\9V_j)$
\beqa
\exp(sL_{(e,e_0)})\exp(\tau L_{(e,e_j)})\9V_j\subset\8{H}_e
\eeqa
and hence
\beqa
\exp((-\pi+s)L_{(e,e_0)})\exp(\tau L_{(e,e_j)})\9V_j\subset\8{H}_{-e}
\eeqa
and therefore $(-\tau,\pi-s))\in I(\9V_j)$ which implies
\beqa
&&b_k(\tau_k,-\pi+s_k)\cdots b_{1}(\tau_1,-\pi+s_k+\cdots +s_1)\in\6B(\8H_{-e})
\vs\vs
&&b_{k+1}^*(-\tau_{k+1},\pi-(s_n+\cdots +s_{k+1}))\cdots 
b_n^*(-\tau_n,\pi-s_n))\in\6B(\8H_e) \ \ .
\eeqa
Keeping in mind that we have  
$$\1b_j(\tau_j)^*=
V_{(e,e_j)}(-\tau_j)\pi(b_j^*)V_{(e,e_j)}(\tau_j)$$
the identity 
\beqa
&&\<V(s_n)\1b_n(\tau_n)\cdots V(s_{k+1})\1b_{k+1}(\tau_{k+1})\Om,
V(s_k)\1b_k(\tau_k)\cdots
\vs\vs
&&\cdots V(s_1)\1b_1(\tau_1)\Om\>
\vs\vs
&=&
\<V(\pi-(s_1+\cdots +s_k))\1b_1(\tau_1)^*V(s_1)\1b_2(\tau_2)^*\cdots 
V(s_{k-1})\1b_k(\tau_k)^*
\Om,
\vs\vs
&&\times \ 
V(\pi-(s_{k+1}+\cdots +s_n))\1b_{k+1}(\tau_{k+1})^*
V(s_{k+1})\1b_{k+2}(\tau_{k+2})^*\cdots 
\vs\vs
&&\cdots V(s_{n-1})\1b_n(\tau_n)^*\Om\>
\eeqa
follows which expresses the KMS condition in the euclidean points.
Finally, a straight forward application of the analysis of \cite{KlLan81}
proves the theorem.
\epr

\section{Proof of Proposition \ref{pro1}}
\label{app3}
The $e_0$-PCT operator $\9J_{(e,e_0)}$ commutes with the 
local one-parameter group $V_{(e,e_1)}$ for $e_0\perp e_1$ 
and it fulfills the relation 
\beqa
\9J_{(e,e_0)}V_{(e,e_0)}(s)\9J_{(e,e_0)}&=&V_{(e,e_0)}(-s)
\eeqa
as easily can be verified. For a time-zero operator 
$b\in\6B(e,\8P_{(e,e_0)})$, for a family of directions 
$\1e=(e_0,\cdots,e_n)$ with
$e_i=e_j$ or $e_i\perp e_j$, $i,j=0,\cdots,n$ and 
for $t=(t_1,\cdots,t_n)\in\7R^{n}$, we introduce an operator 
$\Phi_{(e,\1e)}[t,b]\in\6A(\9W_{(e,e_0)})$, localized 
in the wedge $\9W_{(e,e_0)}$:  
\beqa
\Phi_{(e,\1e)}[t,b]
&:=&
\al_{(e,e_1,t_1)}\cdots\al_{(e,e_n,t_n)}\pi(b) \ \ .
\eeqa
The wedge algebra $\6A(\9W_{(e,e_0)})$ is generated 
by these operators and since the operator 
\beqa
\9J_{(e,e_0)}
\Phi_{(e,\1e)}[t,b]\9J_{(e,e_0)}
&=&
\Phi_{(e,\1e)}[\sgm t,j_{e_0}b] 
\eeqa
is contained in $\6A(\9W_{(e,-e_0)})$, the result follows,
where $(\sgm t)_j=t_j$, if $e_j\perp e_0$, and $(\sgm t)_j=-t_j$, 
if $e_j=e_0$.
\epr

\section{Proof of Theorem \ref{the1}}
\label{app4}
By following the analysis of \cite{KlLan81}, we choose 
a family of operators $b_1,\cdots, b_n$ which are contained 
in the time slice algebra, where 
$b_j\in\6B(\9V_j)$ is localized 
in a convex subset $\9V_j\subset H_{(e,e_0)}$. 
By using the same notations as for the proof of Theorem \ref{the0},
we obtain by putting $V:=V_{(\eta,e,e_0)}$ and $\9J:=\9J_{(e,e_0)}$:
\beqa
&&V(s_k)\1b_k(\tau_k)\cdots V(s_1)\1b_1(\tau_1)\Om
\vs\vs
&=&
V(s_k)\1b_k(\tau_k)V(-s_k)V(s_k+s_{k-1})\1b_{k-1}(\tau_{k-1})
V(-s_k-s_{k-1})\cdots
\vs\vs
&&\cdots V(s_1)\1b_1(\tau_1)\Om
\vs\vs
&=&
V(s_k)\1b_k(\tau_k)V(-s_k)\cdots V(s_1+\cdots+s_k)\1b_1(\tau_1)
\Om
\vs\vs
&=&
\Psi[b_k(\tau_k,s_k)\cdots b_1(\tau_1,s_1+\cdots+s_k)] \ \ .
\eeqa
We compute for $s_1,\cdots,s_k\in\7R_+$ and
$(\tau_i,s_k+\cdots+s_i)\in I(\9V_i)$ for $k\leq i\leq 1$:
\beqa
&&\9J V(s_k)\1b_k(\tau_k)\cdots V(s_1)\1b_1(\tau_1)\Om
\vs\vs
&=&
\Psi
[j_{e_0}(b_k(\tau_k,s_k)\cdots b_1(\tau_1,s_1+\cdots+s_k))]
\vs\vs
&=&
\Psi
[b_1(\sgm_1\tau_1,\pi-(s_1+\cdots+s_k))^*\cdots b_k(\sgm_k\tau_k,\pi-s_k)^*]
\vs\vs
&=&
V(\pi-s_1+\cdots+s_k)\1b_1^*(\sgm_1\tau_1)V(s_1)
\cdots \1b_k^*(\sgm_k\tau_k)^*V(s_k)\Om 
\eeqa
with $\sgm_j=1$ if $e_j\perp e_0$ and $\sgm_j=-1$ if $e_j=e$.
Performing an analytic continuation within the parameter 
$s_1,\cdots s_k$ and $\tau_1,\cdots,\tau_k$ and taking
boundary values at $s_j=\tau_j=0$ yields the relation
(compare \cite{KlLan81} as well as \cite{FrohOstSeil} and \cite{Schl97}) 
\beqa
&&\9J_{(e,e_0)} 
\biggl[\prod_{j=1}^k \exp(t_j\1B_{(e,e_j)})\1b_j\exp(-t_j\1B_{(e,e_j)})\biggr]
\Om
\vs\vs
&=&
V_{(\eta,e,e_0)}(\pi)
\biggl[\prod_{j=1}^k\exp(t_j\1B_{(e,e_j)})\1b_j\exp(-t_j\1B_{(e,e_j)})\biggr]^*
\Om 
\eeqa
which implies that the Tomita operator is 
\beqa
J_{(e,e_0)}\Delta_{(e,e_0)}^{1/2}&=&
\9J_{(e,e_0)}V_{(e,e_0)}(\pi)
\ \ .
\eeqa
Moreover, according to Theorem \ref{the0},
the automorphism group 
\beqa
\al_{(e,e_0)}:t\mapsto \Ad[U(\exp(tB_{(e,e_0)})] 
\eeqa
maps $\6A(\9W_{(e,e_0)})$ into itself and the state 
\beqa
\om_{(e,e_0)}&=&\om|_{\6A(\9W_{(e,e_0)})}
\eeqa
is a KMS state at inverse temperature
$\be=2\pi$ and the theorem follows. 
\epr
\end{appendix}
\newpage


\end{document}